\documentclass[12pt]{article}
\usepackage{graphicx}
\begin{document}

\begin{center}
{\bf Model of World; her cities, languages and countries}

\bigskip
\c{C}a\u{g}lar Tuncay

Department of Physics, Middle East Technical University

06531 Ankara, Turkey

caglart@metu.edu.tr

\bigskip

\end{center}

\textbf{Abstract}

The time evolution of Earth with her cities, languages and countries is
considered in terms of the multiplicative noise \cite{one} and the
fragmentation%\cite{two} 
 processes, where the related families, size
distributions, lifetimes, bilinguals, etc. are studied. %In \cite{three} 
Earlier we treated the cities and the languages differently
(and as connected; languages split since cities split, etc.). Hence,
two distributions are obtained in the same computation at the same
time. The same approach is followed here and Pareto-Zipf law for the
distribution of the cities, log-normal for the languages, decreasing
exponential for the city families (countries) in the rank order over
population, and power law --2 for the language families over the
number of languages in rank order are obtained theoretically in
this combination  for the
first time (up to our knowledge) in the literature; all of which are in
good agreement with the present empirical data.% \cite{four,five,six}.


%PACS numbers:

\section{Introduction}
On Earth there are $235$ countries (states, where the provinces are
disregarded), which have totally about few million cities, where
more than 6 billion humans live and speak about $7000$ languages, at
present. The whole life goes on over an area of some $510$
million km$^2$. Figure 1 is plotted utilizing the present empirical
data \cite{six} for the relative population of the countries (divided
by the world population, thick line) in rank order; i.e., the most
populous country (China) has the rank one, the next populous country
(India) has the rank two, and the third populous country (USA) has
the rank three, etc., where the corresponding relative area of the
countries (divided by the area of the Earth) is designated by the
plot in the thin line. The area plot depicts fluctuation; yet, it
roughly follows the population line. The inset (Figure 1) is for the
population density in capita per km$^2$, which shows that the
population density is almost constant (about $18.5$ capita per
km$^2$) over the countries.

   Our essential aim in the present contribution is to show that the time
evolution of the world with her cities, languages and countries,
etc., might be governed by two opposite processes: random
multiplicative noise for growth in size and fragmentation for spread
in number and extinction. Secondly, we aim at obtaining a wide
panorama for the world (cities, languages, countries and their
distributions, lifetimes, etc.) in terms of a single simulation,
where the related results are obtained at the same time.  The model is 
developed in \cite{three}, where the cities and the languages are treated
differently and as connected; languages split since cities split,
etc. (For a quantitative method for the formation of the languages,
please see references in \cite{three}.) Results for the size
distribution functions, the probability distribution functions (PDF)
and various other functions for both the cities and the languages
are found to be in good agreement with the empirical data. Yet, the
results for the language families (in \cite{three}) are considerably
far from reality.

   In the present work, our focus is on the families; here the size
distributions of the families for both the cities (countries) and
the languages are given besides their distributions over the number
of their members, etc. In \cite{three}, the number of the language
families was not changing in time and the city families were not
considered at all. Here, both the offspring cities and the languages
may create new families; in other words, the current families may
fragment as explained in Sec. 3.3. Secondly, we apply random
punctuation for the cities and random change of the languages, which
was not followed in \cite{three}. We now also study bilinguals.\cite{seven} 
Thus, we present here a richer panorama of the world,
where all of the results are obtained in the same simulation, at
the same time for many parameters.

   The following section is the model, and the next one is the applications and
results. The last section is devoted for discussion and conclusion.
Appendix is a brief description of the model, which is given
extensively in \cite{three}.

\section{Model:}
This section is the definition (2.1.) and a brief review (2.2.-2.4.)
of the model, where also introduced are the meaning of the relevant
concepts and the parameters, with the symbols in capital letters for
the cities and those in lower case for the languages. The subscript
fam is used for the families.

   The initial world has $M(0)$ ancestors for the cities $(I)$ and $m(0)$ ancestors for the languages
$(i)$, with $M(0)\neq m(0)$. Each city has a random size $(P_I(0))$
and she speaks one of the initial languages, which is selected
randomly. So, $P_I(0)$ is the population of each ancestor city, and
$p_i(0)$ is the number of people speaking each ancestor language.
(It is clear that the total number of the citizens and the speakers
is the same and it is equal to the initial world population.)

\subsection{Definition:}
Populations of the cities grow in time $t$, with a random rate $R_I \le R$,
where $R$ is universal within a random multiplicative noise process,

$$P_I(t) = (1 + R_I)P_I(t - 1)   .                       \eqno(1)$$
As the initial cities grow in population the initial languages grow
in size $(p_i(t))$, where the cities (and consequently, the
languages) fragment in the meantime. If a random number (between 0
and 1; defined differently at each time step $t$) 
for a city is larger than some
$G$ close to $1$, then the city becomes extinct (random elimination,
punctuation); otherwise, if it is smaller than some small $H$, the
city splits after growing, with the splitting ratio (fragmentation,
mutation factor) $S$: If the current number of habitants of a city
$I$ is $P_I(t)$, $SP_I(t)$ many members form another population
and $(1-S)P_I(t)$ many survive within the same city. The number of
the cities $M(t)$ increases by one if one city splits; if any two of
them split at $t$, then $M(t)$ increases by two, etc. When a city is
generated she speaks with probability $h_f$ a new language, with
probability $h_s$ a randomly selected current language, and with the
remaining probability $1 - h_f - h_s$ the old language (of the
mother city).

\subsection{Lifetimes for cities or languages:}
Lifetime is the difference between the number of the time step at
which a city or a language is generated and that one at which the
given agent became extinct. The agent becomes extinct if its
size becomes less then unity in terms of fragmentation or if it
is randomly eliminated (with $G<1$). If all the cities which were
speaking a given language are eliminated (by any means), then we
consider the given language(s) as eliminated. And, if all the
members of a family become extinct (by any means), then we consider
the family as extinct. The age of a living agent (at the present) is
considered as the time passed from the time of their formation up to
now.

\subsection{Family trees for the cities or the languages:}
We construct the family trees for the cities and the languages as follows:
We assume that the initial cities and the initial languages have
different families; i.e., we have $F(0)$ many city families and
$f(0)$ many language families at t=0. We label each city by these
numbers, i.e. the city family number and the language family
number, which may not be the same later (for example, due to long
and mass immigration, as in reality). In this manner, we are
able to compute the number of the members of each family, as well as
their sizes at the present time, etc. (The given labels may be also
utilized to trace the generation level of the offspring agents.) It
is obvious that the unification (merging) of the cities or the
languages are kept out of the present scope.

\subsection{ Bilinguals:}
Some citizens of a given city (country) may select another language
(other than the common or official language of the home city, home
country, i.e., the mother tongue) to speak, where several reasons
may be decisive.  We consider here the size distribution of the 
second languages (bilinguals \cite{seven}), where we assume that an adult 
(speaking a language $i$ as a mother tongue) selects one of the current 
languages if this language $k$ is bigger than the mother language,
$p_k > p_i$. Then

$$p'(t)_k \propto p(t)_i(p(t)_k - p(t)_i)\lambda r'_i   ,  \eqno(2)$$
where, $p(t)_i < p(t)_k$ and the prime denotes the second
language. In Eq. (2) $r'_i$ is a random number which is uniformly
distributed between zero and one. So; $0\le \lambda r'_i
<\lambda$ for a given $\lambda$, which is proportional to the
percentage (up to randomness) of the population of the language
$i$ the speakers of which select $k$ as the second language, and
$\lambda$ is taken as universal. It is obvious that, $\lambda$ has the
unit per capita (person) and as the size difference $(p(t)_k -
p(t)_i)$ increases, the language $k$ becomes more favorite and the
related percentage $((p(t)_k - p(t)_i)\lambda r'_i)$ increases.

\section{Applications and results:}

The parameters for the rates of growth (Sect. 2, with the symbols in
capital letters for the cities and these in lateral ones for the
languages) have units involving time: here, the number of the
interaction tours may be chosen as arbitrary (without following
historical time, since we do not have historical data to match
with); and the parameters (with units) may be refined accordingly.
Yet, our initial conditions (with the given initial parameters for
ancestors) may be considered as corresponding to some $10,000$ years
ago from now. So the unit for our time steps may be taken as (about)
$5$ years, since we consider $2,000$ time steps for the evolution.
After some period of the evolution in time we (reaching the present)
stop the computation and calculate PDF for size, and for some other
functions such as extinction frequency, lifetime, etc. (for the
cities or the languages and their families, etc.).
 
   Empirical criteria for our results are: i) The number of the living
cities (towns, villages, etc.) and that of the living languages may
be different; but, total size for the present time must be the same
for both cases (and also for the families of the cities or the
languages), where the mentioned size is the world population (Eqn.
(2)). ii) World population increases exponentially with
time.\cite{four,five,six} iii) At present, the biggest language (Mandarin
Chinese) is used by about $1.025$ billion people and world
population (as, the prediction made by United Nations) is $6.5$
billion in $2005$, (and will be about $10$ billion in $2050$)
\cite{four,five,six}; so the ratio of the size for the biggest language
to (the total size, i.e.,) world population must be (about) $1:6.5$.
iv) Size distribution for the present time must be power law --1 for
the cities (Pareto-Zipf law), and this may be considered as slightly
asymmetric log-normal for the languages. We first consider the
cities (Sect. 3.1), later we study the languages (Sect. 3.2), with
the lifetimes, etc., in all; and the families are considered finally
(Sect. 3.3).
 
\subsection{Cities:}

The initial world population ($W(0)$) is about $M(0)P_I(0)/2$, since
the average of uniform random numbers between zero and unity is 1/2.
Thus, we assume power law zero for the initial distribution of the
cities or the languages over size.
 
   We tried many smooth (Gauss, exponential, etc.) initial distributions
(not shown); and, all of them underwent similar time evolutions
within 2,000 time steps, under the present processes of the random
multiplication for growth, and random fragmentation for spread and
origination and extinction, where we utilized also various
combinations of the parameters $H$ and $G$. We tried also delta
distribution, which is equivalent to assuming a single ancestor, for
the initial case; it also evolved into a power law about --1 (with
different set of parameters, not shown) in time. Since we do not
have real data for the initial time, we tried several parameters for
$M(0)$ (=1,000; 300; 50, etc.) and for $P_I(0)$ (=1,000; 500) at
$t=0$. In all of them, it is observed that the city distribution at
present (Pareto-Zipf law) is independent of the initial (probable)
distributions, disregarding some extra ordinary ones. Please note
that, similar results may be obtained (not shown) for $M(0)=1$,
i.e., single ancestor. 

{\it Evolution:} As $t$ increases, the cities start
to be organized; and within about 200 time steps, we have a picture
of the current world which is similar to the present world, where
the distribution of the cities over population is considerably far
from randomness. With time, the number of the cities ($M(t)$) and
the population of the world ($W(t)$) increases exponentially with
different exponents.\cite{three} Please note that these simulations 
have about two
million cities and the world population comes out as about 4.5
billion at $t=2000$ (present time, the year 2000), for $M(0)=1000$,
$P(0)\le 1000$, $R=0.0075$, $H=0.006$, $G=0.9992$ and $S=0.5$.
With another set of the parameters; for $R=0.0073$, $H=0.004$ and
$G=1$ (keeping other parameters same as before) we have about 450
thousand ($M(2000)$) cities with $6.7$ billion total citizens
($W(2000)$), etc.
 
   In Figure 2 the plots in circles (open ones for $t=320$ and solid ones
for $t=2000$) represent the time evolution of size distribution of
the cities (PDF), all of which split and grow by the same parameters
(Set 1), where $G=0.9992$. Thus, we have abrupt (punctuated)
elimination of the cities here, which is not followed in
\cite{three}; yet the results are not much different, because the
punctuation we applied here is light (low). This means that it is
not strong enough to disturb the running processes, where the
negative effect of the (light) punctuation (in decreasing the
numbers) is diminished by the positive effect of the fragmentation (in
increasing the numbers). Please note that, in Fig.2 the (dashed)
arrow has the slope --1, which indicates the (empirical) Pareto-Zipf
law for the cities. 

Furthermore, we observe that, as the initial
cities spread in number by fragmentation, the initial random
distribution turns out to be log-normal for intermediate times (as
the parabolic fit indicates, for $t=320$ for example) which becomes
a power law --1 (at tail, i.e., for big sizes) for the present time.
The inset (right) in Fig. 2 is the distribution of the world
population ($P$) at $t=320$ (dashed line) and $t=2000$ (solid line)
over the cities ($C$), which are in rank order along the horizontal
axis. Please note that, in the figure and in the inset, axes are
logarithmic. It may be observed in the plots in the inset (Fig. 2,
right) that, the world population (along the vertical axis, $P$)
increases slightly more rapidly than the number of the cities (along
the horizontal axis, $C$). So, Fig. 2 may be considered as the
summary for history of the evolution of the cities (or the
languages, see Section 3.2.), where two opposite physical processes
underline the evolution; the random multiplicative noise and the
fragmentation.

   {\it Lifetimes:} We obtain the time distribution of the cities (lifetime
for extinct cities and ages of the livings ones, not shown here) as
decreasing exponentials (disregarding the cases for small number of
ancestors and high punctuation) as given in the related figures in
\cite{three}. Simple probability (density) functions for the
lifetimes are also exponential (not shown), which means that the
cities occupy the time distribution plots in exponential order; more
cities for small $t$, and fewer cities for big $t$, for a given number
of time steps in all.

\subsection{Languages:}

We guess that there were many simple languages (composed of some
fewer and simple words and rules), which were spoken by numerous
small human groups (families, tribes, etc.) at the very beginning.
And, as people came together in towns, these primary languages might
have united. Yet, we predict that the initial world is not (much)
relevant for the present size configuration of the languages (as
well as in the case for cities; see, Sect. 3.1). Moreover, we may
obtain similar target configurations for different evolution
parameters (not shown). Within the present approach, the ancestor
cities and the ancestor languages are associated randomly; since,
the languages with their words, grammatical rules, etc. might have
been formed randomly (\cite{three}, and references therein); the
societies grew and fragmented randomly (as mentioned in Sect. 2);
new cities randomly formed new languages or changed their language
and selected a new one randomly. We predict that, the index $i$
(roughly) decreases as $I$ increases for small $I$ (not shown). We
predict also the distribution of the present languages over the
present cities, where we have power law minus unity (not shown). It
may be worthwhile to remark that, younger cities prefer younger
languages; which means also that the new cities (or the new
countries which are composed of the new cities) emerge mostly with
new languages. Secondly, as $t$ increases the indices $I$ and $i$
increase, and the plot of $I$ versus $i$ extends upward and
moves rightward, since the number of the current languages ($m(t)$)
and the number of cities, which speak a given language, increase (as
a result of the fragmentation of the cities). 

Furthermore, we
compute the number (abundance) of the speakers for the present
languages ($p_i(t)$, in Eq. (2)) (not shown), where we have few
thousand ($m(t=2000)=7587$) living languages. Within this
distribution of the present languages over the speakers, we predict
power law minus unity (not shown). It may be worthwhile to remark that
older languages have more speakers; and in reality (Mandarin)
Chinese, Indian, etc., are big and old languages. For example, we
have about one billion people speaking the language number 1, which
is one of the oldest languages of the world; and less people
speaking the language number 2, etc.

  In Figure 2, we display the PDF for the size distributions of the
languages at $t=320$ (historical, open squares) and $t=2000$
(present, solid squares), where the number of the ancestor languages
($m(0)$) is 300. We plotted several similar curves for $m(0)=1$
i.e., for the case where only one ancestor language is spoken in
each ancestor city and obtained similar results (not shown).
Splitting rate and splitting ratio for languages are not defined
here, since languages split as a result of splitting of the cities;
and the splitting ratio of the splitting language comes out as the
ratio of the population of the new city (which creates a new
language) to the total population of the cities which speak the
fragmented language. Please note that, in the plots (Fig.2) for the
languages at the present time (solid squares for $t=2000$) we have
slightly asymmetric Gauss for big sizes as the parabolic fit (dashed
line) indicates; and we have an enhancement for the small languages
in agreement with reality \cite{eight}. Fig. 2 may be considered as
the summary for history of the evolution of the cities and the
languages.

   We think that, the (random) elimination of the languages (with all of its speakers) is not
realistic (excluding the small languages with small number of
speakers), and it is not recorded in the history for the recent
times. On the other hand, changing (replacing) a language by another
one may be realistic. And, in case of random (light) elimination
(i.e., changing the language with a current one), the fragmentation
rate may accordingly be increased to obtain the empirical data for
the number of the languages at the present. In other words, the
number of the languages increases by $h_f$ and decreases by $h_s$,
which may be considered as punctuation for the languages with $1 -
h_s< g$. In Fig. 2 (and in other related ones) we utilized $h_f
=0.0013$ and $h_s=0.0002$.

   Lifetimes for the languages and the related probability
densities are decreasing exponentially (as for the cities); which
means that many languages (cities) become extinct soon after they emerge
and the remaining ones live long (not shown) as in reality. Please
see Fig. 2 in \cite{eight} for the related empirical data. The
(negative) exponent of the present decay is about $0.0007$ per time
step for $2000$ time steps.

\subsection{Families of the cities (countries) or the languages and the bilinguals:}
We obviously do not know how the city (language) families \cite{eightplus} are
distributed over the cities (languages) initially; since, we do not
have any historical record about the issue. Yet, we predicted that
the initial conditions for the cities (languages) are almost
irrelevant for the present results. And we considered several
initial conditions for the city families and the language families,
which are discussed in \cite{eight} to some extent in empirical
terms.
 
   We think that, the number of the city families and the language families were roughly the
same, (yet, the number of the cities and the languages might be
different) initially; and we take $F(0)=30$, $f(0)=18$  (for
$M(0)=1000$ and $m(0)=300)$.
   Figure 3 and Figure 4 are for the city families and
the language families, with $H_{\rm fam-f} =0.0005$, $H_{\rm fam-s} =0.0003$
$(G_{\rm fam}=1 - H_{\rm fam-s})$ and $h_{\rm fam-f} =0.0004, h_{\rm fam-s} =0.0001 \; (g_{\rm fam}=1 -
h_{\rm fam-s})$, all respectively; where, other parameters are as before.
Figs. 3 and 4 may be considered as in good agreement with the
empirical plots in Fig. 1 (and in refs. \cite{four,five,six}). Please
note that, the families evolve in time here (with the parameters
given for the related fragmentation in this paragraph), which is not
considered in \cite{three}.
   For the bilinguals we assume that a small fraction $\lambda$ of the
citizens selects a second language out of the bigger ones and the
introduced probability increases with the difference in sizes.
Figure 5 is the PDF for the distribution of the bilinguals over the
relative population (to the total), where Eq. (3) is utilized with
the present languages (Figs. 2 and 4) for  $\lambda=0.01$. We
observe in Fig. 5 that few big languages are favored as the second language 
by the majority (about 90 \%) of the speakers, with the given $\lambda$ .
We think that selecting big languages as second languages may help
increase the sizes of the big languages.

\section{Discussion and conclusion:}

Starting with random initial conditions and utilizing many
parameters in two random processes (the multiplicative noise for
growth and the fragmentation for generation and extinction of the
cities or the languages) for the evolution, we obtained several
regularities (for size and time distributions, etc.) within the
results; all of which may be considered as in good agreement with
the empirical data.

   We predict that the results are (almost)
independent of the initial conditions, disregarding some extra
ordinary ones. Furthermore, punctuation (besides fragmentation)
eliminates the ancestors, with time. For $G\neq1$ ($G\approx1$), we
need longer time to mimic target configuration (if other parameters
are kept the same as before), where new generated cities or
languages may be inserted, in terms of fragmentation, provided
$G_{\rm critical}\le G$, and $1\le G+H$.

   Many cities or languages become extinct in their youth, and
less become extinct as they become old. In other words,
languages or cities become extinct either with short lifetime (soon
after their generation), or they hardly become extinct later and
live long (which may be considered as a kind of natural selection).
We consider the mentioned result (which is observed in reality
\cite{eight})as an important prediction of the present model and we
had obtained similar results for the evolution biological species
\cite{nine}, which may be coming out because of the present random
multiplicative noise and fragmentation processes.

   It might be argued (objected, by the reader) that,
there are many parameters in the model. Each of them is needed for
some measure of the related evolution in reality. Secondly, as we
predict that the initial conditions are (almost) irrelevant for the
present results and many parameters (for $t=0$) may be ignored.
Thirdly, the most important parameter in the model is $R$ (the rate
for population growth), and $H$ is related to $R$ implicitly, since
we need more cities to be established (per time) as the world
population increases. The punctuation parameter $G$ may also be
considered as a parameter dependent (implicitly) on $R$; since the
probability for the emergence and spread of wars, illnesses, etc.
increases, as world population increases. The rates for the
languages ($h$, $g$, etc.) may also depend (implicitly) on $R$;
since more new languages (per time) are needed with the increasing
world population, etc. The rates for the families certainly depend
on the number of their current members, the rate of which may
(ultimately) be controlled by $R$. Speaking geometrically, the area
under each plot for the size distribution of the cities, languages,
language families, city families (countries) must always equal to
the world population at any time $t$; and this constraint constructs
the bridge for the given implicit dependence of the rate parameters
(for the considered functions) on $R$.

   As a final remark we claim that the
original model may be useful to predict also the historical size
distribution of the cities: We predict that the initial distribution
of the cities over the population becomes parabolic for some
intermediate time $t$ ($<2000$) in log-log scale and it turns to be
power law --1 as time goes on (i.e., for the present time; $t=2000$).
The mentioned distribution may be checked within the archaeological
data (as a subject of a potential field of science; namely, physical
history) for the ancient cities (towns); where, the time evolution
of the mentioned distribution into power law --1 may also be
considered.

\section{APPENDIX}

In the present model, we have (with the symbols in capital letters
for the cities and those in lower case for the languages; and the
sub index fam is used for the families) $M(0)$ ancestors for the
cities $(I)$ and $m(0)$ ancestors for the languages $(i)$, with
$M(0) \neq m(0)$. Each city has a random size ($P_I(0)$) and she
speaks one of the initial languages, which is selected randomly. So,
($P_I(0)$) is the population of each ancestor city, and $p_i(0)$ is
the number of people speaking each ancestor language. It is clear
that the total number of the citizens and the speakers is same (for
any $t$) and it is equal to the current world population;

$$W(t) = \sum^{M(t)}_{I=1} Pi_I(t) = \sum^{m(t)}_{i=1} p_i(t) . \eqno(A)$$ 

The cities have fixed growth rates ($R_i$), which are distributed
randomly over the ancestors and they (and, these for the offspring;
where the offspring carry the same growth rate as their ancestors)
are not changed later; yet, the maximum value ($R$) for the growth
rates is constant for all of the cities (so is for world).
Furthermore, we have $F(0)$ initial city families and $f(0)$ initial
language families, with $F(0)\ne f(0)$. Please note that, all of the
introduced parameters are about physical quantities, which represent
several situations in reality. 

The time evolution of the cities (or
the languages and their families) is considered in terms of two
random processes, the multiplicative random noise \cite{one} and the
random fragmentation  \cite{two}, which are coupled; here, the
cities or the languages (and their families) are taken as a whole
and the individuals are ignored. The cities grow in number by
splitting (with constant ratio $S=1/2$) where, the fragmentation
rate is $H$; and, the languages, the city families and the language
families follow them accordingly, with various fragmentation rates:
If a new city forms a new language ($h_f$) then it means that, the
language of the home city is fragmented; here, the splitting ratio
($S$) is the ratio of the population of this new city to the total
population of the cities which speak the old language. It is obvious
that $h_f$ is small ($h_f\approx0$); yet, many new languages may
emerge at each time step, since many new cities emerge in the mean
time, and $h_f$ becomes important. On the other hand, a new (and an
old) city may change her language and select one of the current
languages as the new one (with $h_s\approx0$, for all), where
colonization may take place or teachers may teach the new language
 \cite{three}, etc. In this case, size of the old (new) language decreases
(increases) by the population of the new city. The language which is
spoken by many cities has a higher chance for being selected by a new
city; and so, big languages are favored in case of selecting a new
language.

   We consider the countries (city families) as follows: When a city is
newly generated she establishes a new country (state, as we know
many historical examples where each city was a state (city-state)
and many new countries started with a new city) with probability
$H_{\rm fam-f}$; with probability $H_{\rm fam-s}$ she is colonized (i.e., changes
country); and, with the remaining probability $1 - H_{\rm fam-f} -
H_{\rm fam-s}$ she continues to survive within the home country. It is
obvious that when a city (or a group of cities, due to the present
randomness) starts a new country, it means that the old country is
fragmented. Secondly, not only the newly generated cities but also
the old ones may be colonized. The countries with all of her cities
may also be colonized (conquered) as we know from many examples in
history. Similar treatment may cover the language families with the
parameters $h_{\rm fam-f}$, $h_{\rm fam-s}$ and $1 - h_{\rm fam-f} - h_{\rm fam-s}$ (the
probability for starting a new language family, for changing the
language (and so the language family, while surviving within the
home country, i.e., being culturally colonized) and for continuing
to speak a language which belongs to the home language family;
respectively).

   Please note that, the fragmentation causes new agents
to emerge (birth), and at the same time it drives them to extinction
in terms of splitting, and any agent with a member less than unity
is considered as extinct. The number of the cities increases,
decreases, or fluctuates about $M(0)$ for relatively big numbers for
$H$ (high fragmentation) and $G$ (low elimination), for small
numbers for $H$ (low fragmentation) and $G$ (high elimination), and
for $H + G = 1$ (equal fragmentation and elimination), respectively;
out of which we regard only the first case, where we have (for
$1<H+G$) an increase in the number of cities, and we disregard the
others. We try several numbers of the ancestors $M(0)$, with sizes
$P_i(0)$, where we assign new random growth rates for the new
cities, which are not changed later, as well as the growth rates for
ancestors are kept as same through the time evolution.

It is obvious that $H=1=G$ gives the gradual evolution for the cities, where we
have regular fragmentation with $H$ (and with some $S$) at each
time step $t$. This case is kept out of the present scope, because
we consider it as (historically) unrealistic.

   It may be worthwhile to stress that elimination (punctuation, $G<1$) plays a role which is
opposite to that of fragmentation ($H$) and growth ($R$) in
evolution; here, $H$ and $R$ develop the evolution forward, and $G$
recedes. So the present competition turns out to be the one between
$H$ and $R$, and $G$, where two criteria are crucial: For a given
number of time steps, $R$, and $M(0)$, etc., there is a critical
value for $G$; where, for $G_{\rm critical} <G \cong 1$ cities survive, and
for smaller values of $G$ (i.e., if $G\cong G_{\rm critical}$) cities may
become extinct totally. (For similar cases in the competition
between species in biology, one may see [10].) Secondly, sum of H
and $G$ is a decisive parameter for the evolution: If for a given
$G$ (with $G_{\rm critical}  <G$), $H+G=1$, then the number of cities does
not increase and does not decrease, but oscillates about $M(0)$,
since (almost) the same amount of cities emerges (by $H$) and
becomes extinct (by $G$) at each time step, and we have intermediate
elimination. On the other hand, if $H+G<1$, then the cities decrease
in number with time and we have high (strong, heavy) elimination.
Only for $1<H+G$ (with $G\neq1$) we have low (weak, light)
elimination of cities, where the number increases (yet, slowly with
respect to the case for $G=1$). In summary, only light punctuation
of the cities may be historically real, and it does not affect the
evolution and size distribution of the cities, as we observed in
many runs (not shown), where we increase the fragmentation ($H$) and
population growth rate ($R$) to compensate the negative effect of
punctuation on the number of cities and world population,
respectively. Yet, the ancestor cities i.e., those at age of
$t$ at any time $t$, decay more quickly in time as (punctuation
increases) $G$ decreases (since the generated cities may be
substituted by new generated ones after elimination; but the
ancestor ones can not be re-built.) It is obvious that punctuation
of a city (together with all of the citizens) is realistic as many
(regrettable) examples occurred during many wars.

\begin{figure}[hbt]
\begin{center}
\includegraphics[scale=0.99]{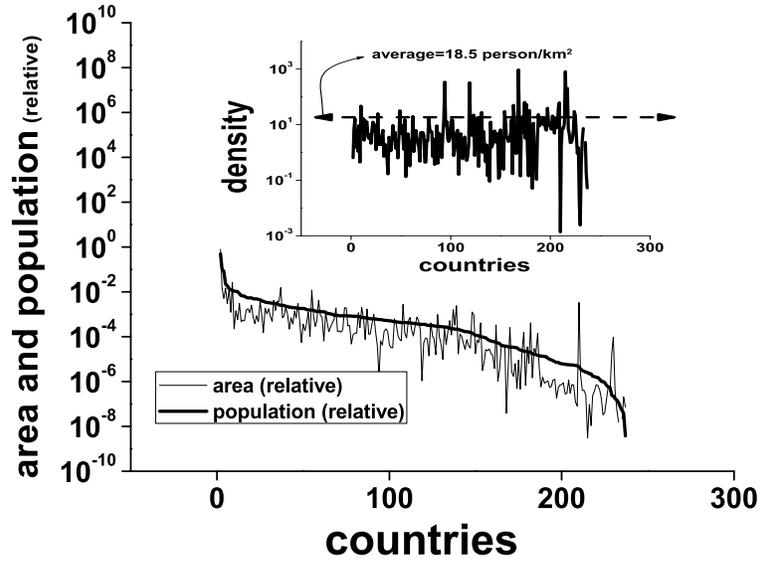}
\end{center}
\caption{ The population (thick curve; in rank order along the
horizontal axis, which is linear) and the area (fluctuating thin
curve) of the present countries. The inset is for the population
density per km$^2$, where the two sided arrow designates the average
which is about $18.5$ people per km$^2$. Please note that, many (about
$200$) countries have the population density between $0.1$ and $1$
capita per km$^2$. (Empirical data of  \cite{six} is utilized to
produce the plots.) }
\end{figure}

\begin{figure}[hbt]
\begin{center}
\includegraphics[scale=0.99]{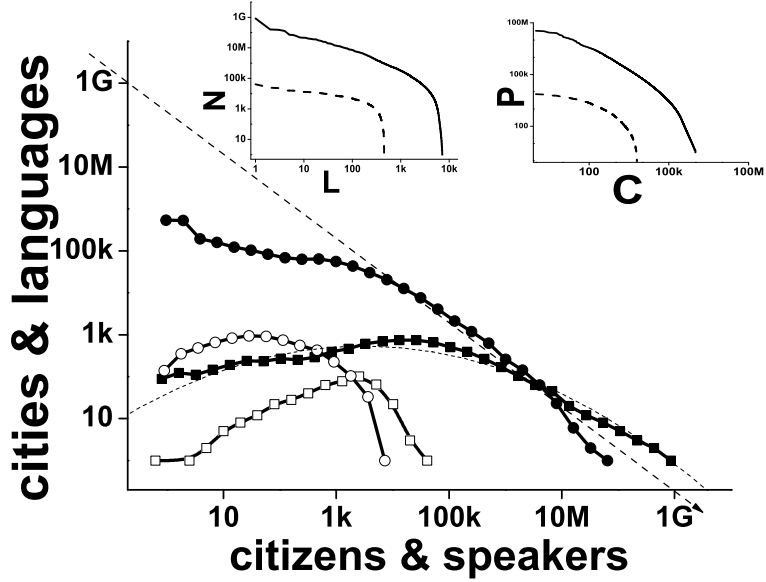}
\end{center}
\caption{ Historical and present size distribution of the cities and
the languages; open circles for the cities at $t=320$, solid circles
for the cities at $t=2000$, open squares for the languages at
$t=320$ and solid squares for the languages at $t=2000$. The dashed
arrow has the slope = --1 (for the cities) and the dashed curve is
parabola (for the languages). We have 6417 historical cities,
$2.046$ million living cities, 504 historical languages and 7882
living languages; historical world population is $1.417$ million and
the present world population is $4.532$ billion (i.e., the areas
under the related lines). The inset (left) is the distribution of
the speakers ($N$) at $t=320$ (dashed line) and at $t=2000$ (solid
line) over the languages ($L$) which are shown in the rank order of
the size along the horizontal axis. The inset at right is the same
as the other inset (left) but for the population ($P$) over the
cities ($C$). Please note that, all of the distributions are
obtained in the same computation at the same time and the axes are
logarithmic (in the figure and in the insets). Parameters are given
in Sect. 3.1 and 3.2. }
\end{figure}

\begin{figure}[hbt]
\begin{center}
\includegraphics[scale=0.99]{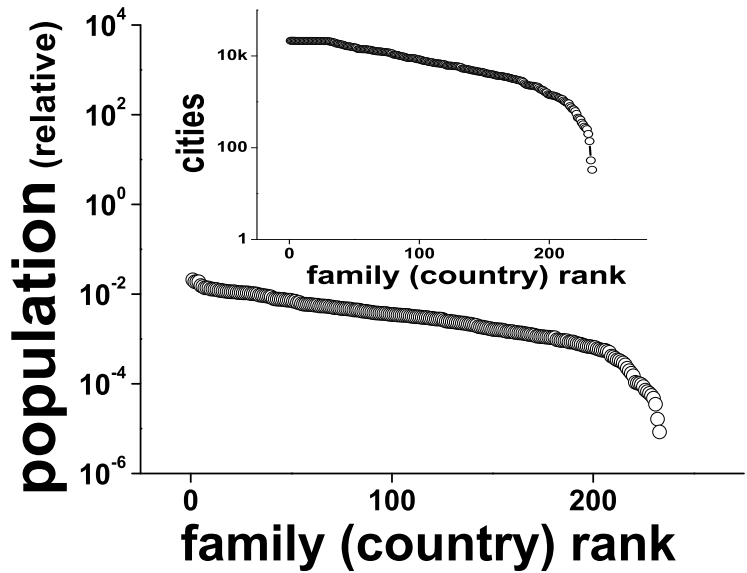}
\end{center}
\caption{ The city families (in rank order along the horizontal
axis, which is linear) and the population, which is relative to the
total (along the vertical axis, which is logarithmic) at the present
time, where we have about 240 families (with 30 families at $t=0$);
other parameters are same as in Fig. 2. Please note that, population
is distributed in decreasing exponential over the city families
(countries) with small exponent, as in Fig.1 (empirical); yet, we
have here a shorter head (on the left), which may be due to the fact
that in Fig. 1 the biggest countries (China, India, USA, etc.) are
displayed with their provinces (where, the countries with provinces
may be considered as the families of the city families). Please note
the similarity for relaxation at the small population end in Fig. 1
and here. The inset is the distribution of the present cities over
the families. }
\end{figure}

\begin{figure}[hbt]
\begin{center}
\includegraphics[scale=0.99]{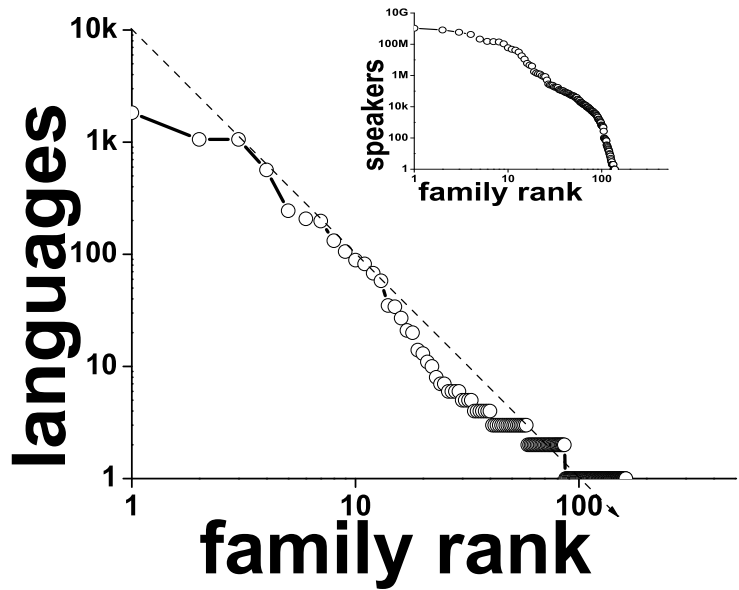}
\end{center}
\caption{ The language families (in rank order along the horizontal
axis, which is linear) and their number of the members (along the
vertical axis, which is logarithmic) at the present time, where we
have about 140 families (with 18 families at $t=0$); other
parameters are same as in Fig. 2. Please note that, we have 7468
languages at $t=2000$, and the dashed arrow has slope --2. The inset
is the size distribution of the present families, where $1.07$
billion people speak the biggest language. }
\end{figure}

\begin{figure}[hbt]
\begin{center}
\includegraphics[scale=0.99]{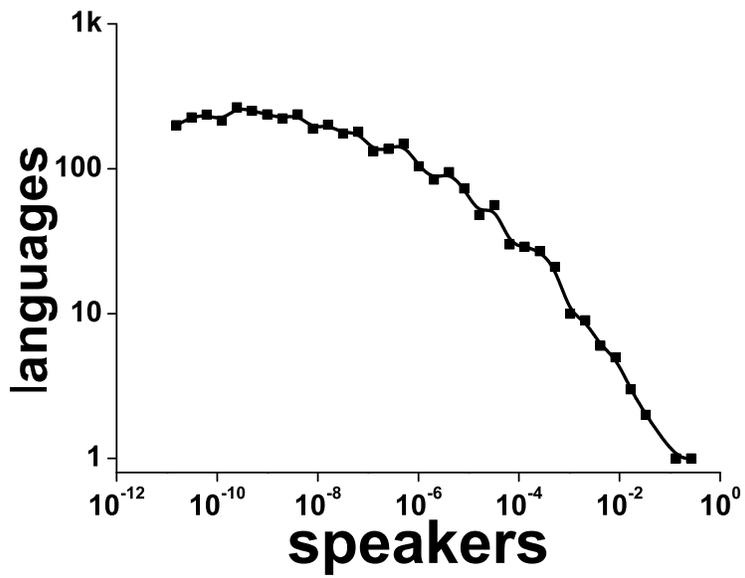}
\end{center}
\caption{ The probability distribution function for the second
languages (bilinguals) over (arbitrary) size, where the bigger
languages (than the mother languages) are favored with the
probability which is taken as proportional to the difference in
sizes.  $\lambda=0.01$; for the other parameters, please see Sect. 3.
}
\end{figure}
\end{document}